\documentclass[
  aps,
  prl,
  reprint,
  superscriptaddress,
  floatfix,
  amsmath,amssymb,
  nofootinbib,
  nobibnotes, 
  nolongbibliography, 
  noeprint 
]{revtex4-2}

\usepackage[utf8]{inputenc}
\usepackage[T1]{fontenc}
\usepackage{graphicx}
\usepackage{siunitx}
\usepackage{physics}
\usepackage{bm}
\usepackage{xurl}
\usepackage[hidelinks]{hyperref}

\begin{document}

\title{Hereditary Effective Atom Theory for Large-Scale Atomic Structure Prediction}

\author{Justin~Tahmassebpur}
\email[Corresponding author: ]{jt577@cornell.edu}
\affiliation{Department of Applied and Engineering Physics, Cornell University, Ithaca, NY, USA}
\author{Chia-Hao Lee}
\affiliation{Department of Applied and Engineering Physics, Cornell University, Ithaca, NY, USA}
\author{Crist\'obal M\'endez}
\affiliation{Department of Applied and Engineering Physics, Cornell University, Ithaca, NY, USA}
\author{Shake Karapetyan}
\affiliation{Department of Applied and Engineering Physics, Cornell University, Ithaca, NY, USA}
\author{Zixiao Shi}
\affiliation{Department of Chemistry and Chemical Biology, Cornell University, Ithaca, NY, USA}
\author{David A. Muller}
\affiliation{Department of Applied and Engineering Physics, Cornell University, Ithaca, NY, USA}
\author{Tom\'as~Arias}
\affiliation{Department of Physics, Cornell University, Ithaca, NY, USA}
\date{\today}

\begin{abstract}
Predicting the atomic structure of large systems requires simultaneous exploration of atomic positions, species, and, generally, the number of atoms. The combinatorial growth of possible arrangements can make this search difficult even for standard global optimization methods. We introduce Hereditary Effective Atom Theory (HEAT), which integrates Effective Atom Theory (EAT) into a genetic algorithm. HEAT combines EAT's continuous relaxation of atomic positions, species, and population with the global search capabilities of genetic algorithms. Central to the method is a crossover scheme that directly superimposes two parent structures and relaxes the combined structure. HEAT outperforms conventional genetic-algorithm baselines in recovering a missing 240-atom H sublattice, predicting a 364-atom edge-dislocation core, and reconstructing a 1150-atom interior region of twisted-bilayer $\mathrm{WSe_2}$ from experimental ptychographic constraints. These results demonstrate how continuous effective-atom relaxation improves the exploration of large combinatorial structure spaces.
\end{abstract}

\maketitle

\section{Introduction}
\label{sec:intro}

Predicting the atomic structure of a large system requires finding a favorable arrangement of atoms among a combinatorial number of possibilities. One must generally search over the positions, species, and quantity of atoms simultaneously. This difficult problem arises in crystals, nanoparticles, interfaces, defects, and disordered materials~\cite{oganov2006crystal,zhu2013surface,revard2016grandcanonical,frolov2013grainboundaryphases,pandey2016diffractioninversion}. Experimental data or other constraints might reduce the search space, but the combinatorial scaling with the number of atoms and species is generally unavoidable. 

Diffusion models offer a complementary approach to generating candidate structures. Models such as CDVAE, DiffCSP, and MatterGen learn from databases of known or calculated crystals and can generate new structures with specified compositions or properties~\cite{xie2022cdvae,jiao2024diffcsp,zeni2025mattergen}. However, success on small unit cells does not establish reliable generation of large structures with defects or disorder. For example, the reported MatterGen base model was trained on crystals containing at most 20 atoms per cell~\cite{zeni2025mattergen}. Moreover, these models are generally black boxes, and their ability to generate low-energy structures containing hundreds or thousands of atoms remains uncertain, particularly when trained primarily on small crystals. Generated structures must still be checked through energy calculations and relaxation, and their quality for systems far outside the training data remains uncertain. This leaves a need for methods that search directly for favorable atomic arrangements in large systems.

Several established methods search directly for low-energy structures. Random structure searching generates and locally relaxes many independent trial structures~\cite{pickard2011airss}, while basin hopping and minima hopping alternate local relaxation with moves that explore other minima~\cite{wales1997basinhopping,goedecker2004minimahopping}. Particle-swarm optimization guides a population of candidates using information from previously found low-energy structures~\cite{wang2010particleswarm}. Genetic algorithms provide a powerful approach to this search by maintaining and evolving a population of candidate structures~\cite{deaven1995molecular,oganov2006crystal,glass2006uspex,curtis2018gator}. Candidates are ranked by their energy or another specified objective, and better candidates have a higher chance of being selected as parents. Crossover creates a child by combining parts of two parents, while mutation introduces changes such as moving, adding, or removing atoms. Each new candidate is locally relaxed before it is evaluated for selection. Repeating these steps over successive generations allows the search to explore many local minima. This approach has been successful in predicting crystal structures, particularly for small and moderately sized unit cells, and developments such as USPEX have substantially extended its capabilities~\cite{oganov2006crystal,glass2006uspex,lyakhov2013uspex}. Frameworks such as FANTASTX also combine discrete-atom genetic operations and local relaxation with experimental constraints to guide atomic-structure searches~\cite{fantastx}.

For large systems, conventional genetic algorithms can struggle to find low-energy structures from random starting configurations. Local relaxation often leaves these candidates trapped in disordered arrangements, and combining two disordered parents can simply produce another poor structure~\cite{lyakhov2010largesystems}. Further improvement requires larger rearrangements that crossover and mutation may take many generations to produce. Introducing order into the initial population and preserving favorable local arrangements can improve convergence, as demonstrated by USPEX~\cite{lyakhov2010largesystems,lyakhov2013uspex}. However, obtaining a useful starting structure can itself be difficult. For example, Yu \emph{et al.} reconstructed a gold nanoparticle containing thousands of atoms, but their search began with crystalline candidates whose orientation was already known experimentally and still proceeded over thousands of generations~\cite{yu2016integratedrefinement}. Such initialization supplies substantial structural information that may be unavailable for an unknown defect or disordered material. This motivates methods that can more readily escape poor local minima without relying on an ordered starting structure.

Effective Atom Theory (EAT) offers a way to address this difficulty~\cite{tahmassebpur2025eat}. It represents each site as a mixture of elements and a null species corresponding to an absent atom, allowing the mixture weights and atomic position to vary continuously during relaxation. A site can gradually change from one element to another, disappear as its weight shifts to null, or reappear as its weight shifts back to a real element. For example, an atom can gradually disappear to allow neighboring atoms to rearrange, avoiding an unfavorable configuration that positional relaxation alone would encounter. Such changes can open paths to lower-energy structures even when the starting configuration is a local minimum with respect to atomic positions. The system is then driven toward a final structure in which each site is either assigned a single real element or removed, leaving only conventional atoms with definite chemical identities.

As originally formulated, EAT is a local optimization method. It follows gradients from a given starting configuration and can still become trapped in a local minimum, especially in a large system. We therefore introduce \emph{Hereditary Effective Atom Theory} (HEAT), a genetic algorithm that uses the effective-atom representation of EAT to enable efficient global search. Central to the method is a novel crossover operation that directly superimposes and subsequently relaxes two parent structures, allowing atoms from either parent to move, change identity, or disappear. The child can thereby retain favorable parts of both parents while removing redundant or unfavorable atoms. Importantly, the method does not require an ordered initial structure, making it more flexible than conventional genetic algorithms.

We demonstrate HEAT by recovering a missing 240-atom H sublattice in Mg(BH$_4$)$_2$, predicting a 364-atom edge-dislocation core in Fe, and reconstructing an interior region of twisted-bilayer $\mathrm{WSe_2}$ containing 1150 atoms from experimental electron ptychography data. Across all three applications, HEAT improves on a conventional genetic-algorithm baseline. In five independent runs of each method for each of the first two benchmarks, HEAT reaches the reference-energy criterion more often within 20 generations. The best HEAT runs recover the reference structures by generations 2 and 4, respectively, with populations of just 30 members. This rapid convergence suggests that, even for hundreds of atoms, EAT's expanded continuous search space provides sufficiently accessible relaxation pathways that only a few rounds of evolutionary selection and recombination are needed to reach the target structures. The improvement is particularly striking for WSe$_2$, our largest test system: after 128 generations, HEAT achieves a substantially lower formation energy and much closer agreement with experimental ptychography data than the conventional genetic algorithm achieves after 352 generations, at approximately equal energy-evaluation budgets. These results demonstrate that combining EAT relaxation with evolutionary search can recover favorable structures containing hundreds to thousands of atoms without requiring ordered initial arrangements.

\section{Results}
\subsection{Hereditary Effective Atom Theory}
\label{sec:methods:HEAT}

\begin{figure*}[tb]
    \centering
    \includegraphics[width=0.99\linewidth]{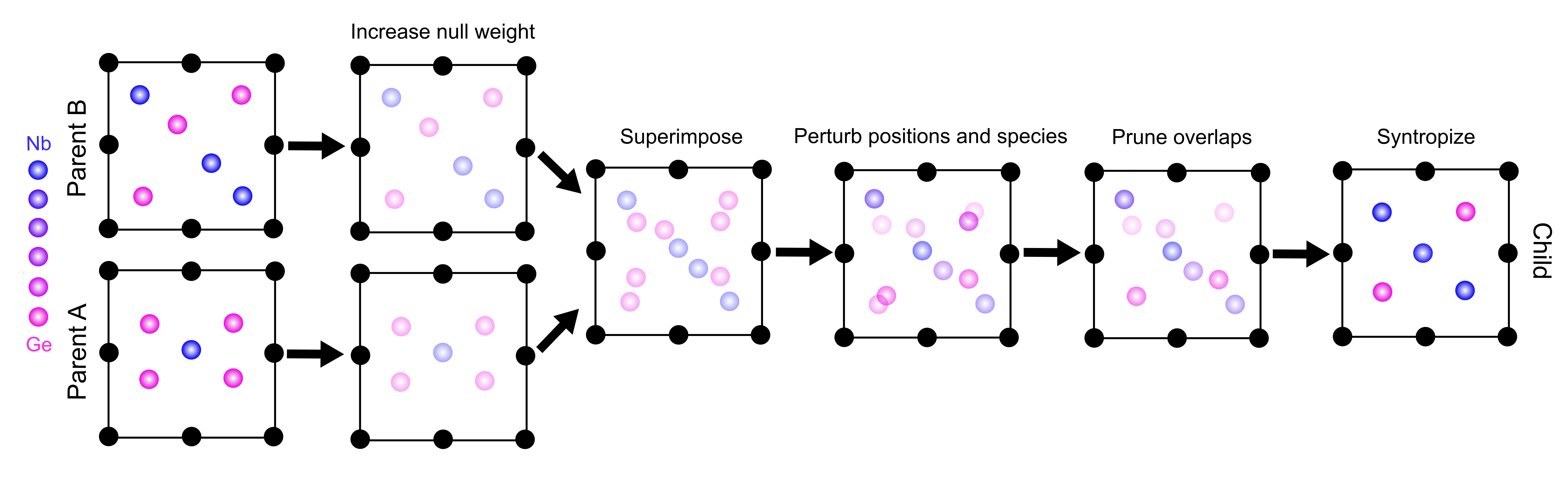}
    \caption{\textbf{Whole-structure superposition crossover in HEAT.}
    Two fully discrete parents are lifted back into effective-atom space by increasing the null weight of every optimizable atom. The two complete parent structures are then spatially superimposed within their common scaffold. Atomic positions and effective-atom weights are perturbed, near-overlapping effective atoms are pruned, and the child is syntropized. Continuous relaxation can retain compatible atoms or motifs from either parent while redundant or unfavorable effective atoms become null. Black atoms denote the fixed scaffold; color denotes relative Nb--Ge character and opacity denotes total non-null occupancy in this illustrative example.}
    \label{fig:genetic_EAT_demo}
\end{figure*}

Hereditary Effective Atom Theory (HEAT) maintains a population of candidate structures and uses EAT to optimize each newly generated member. Generation zero consists entirely of independent, randomly initialized structures, using only the information supplied as input to each reconstruction problem, as described below. No parent structures are required for or involved in this initialization. Each initial structure is optimized through syntropization, which drives effective atoms toward discrete elemental or null states. A final rounding step resolves any remaining effective atoms, and null atoms are removed. The remaining real atoms undergo a final positional relaxation, yielding a discrete candidate whose objective value determines its fitness.

Each subsequent generation is constructed from three groups: newly initialized random immigrants that introduce diversity, elite structures retained from the previous generation, and children produced by crossover between members of the previous generation. To create a child, two parents are chosen from the previous generation by tournament selection~\cite{miller1995tournament,greenstein2023bestpractices}. A randomly sampled group of candidates forms the first tournament, and its lowest-score (best objective value) member becomes the first parent. A second tournament is then drawn after excluding that parent, and its lowest-score member becomes the second parent. This procedure favors low-score structures while allowing a range of candidates to contribute offspring. This parent selection followed by crossover is repeated until the requisite number of children has been generated.

The crossover proceeds through the steps shown in Fig.~\ref{fig:genetic_EAT_demo}. First, we halve the real-species weights of every optimizable atom in both parents and increase the null weights accordingly. The two structures are then spatially superimposed within their common cell. A fixed scaffold, such as atoms at the boundary of the search region, provides a common spatial reference. More generally, any constraint that approximately fixes the positions and orientations of the candidate structures can serve this purpose. Next, we add small perturbations to positions and species weights to introduce variation, and prune effective atoms that lie too close together to avoid severe atomic overlaps. EAT then optimizes the combined structure, allowing atomic positions, identities, and occupancies to adjust together. During the syntropization phase, compatible features from either parent can survive, while redundant or unfavorable effective atoms can disappear into null. Then, as with the initialization procedure described above, hard rounding, removal of null atoms, and final positional relaxation produce the discrete child. Only the final objective value of this discrete structure is used for selection; fractional occupancies do not enter any comparison scores.

For the controlled H-sublattice and edge-dislocation comparisons, HEAT and the conventional GA use the same random, discrete initialization process for generation zero and random immigrants, the same population size and elite/random/crossover fractions within each benchmark, and the same final local-relaxation criterion. Their principal differences are the representation and local optimization used before final scoring, and the crossover and mutation operations. HEAT uses effective-atom relaxation and whole-structure superposition; the conventional GA uses hard atoms, cut-and-splice crossover, and explicit mutations. The ptychographic comparison also includes differences in initialization, as described below. All searches use fixed unit-cell vectors, and any supplied scaffold atoms remain fixed. Full details and run parameters are given in Methods.

Unless otherwise stated, all energies and forces are evaluated using the UMA-S-1.1 model with the \texttt{omat} task in FAIRChem~\cite{wood2026uma}. UMA provides an efficient approximation to DFT energies and forces, with strong accuracy across diverse materials benchmarks~\cite{wood2026uma}. We use UMA rather than DFT because the large system sizes and numerous energy and force evaluations required by these searches make direct DFT calculations prohibitively expensive. EAT optimizations are performed using the machine-learning interatomic potential (MLIP) version of EAT~\cite{tahmassebpur2025eat}.

\subsection{Sublattice prediction}
\label{sec:sublatt}

\begin{figure*}[t]
    \centering
    \includegraphics[width=0.99\linewidth]{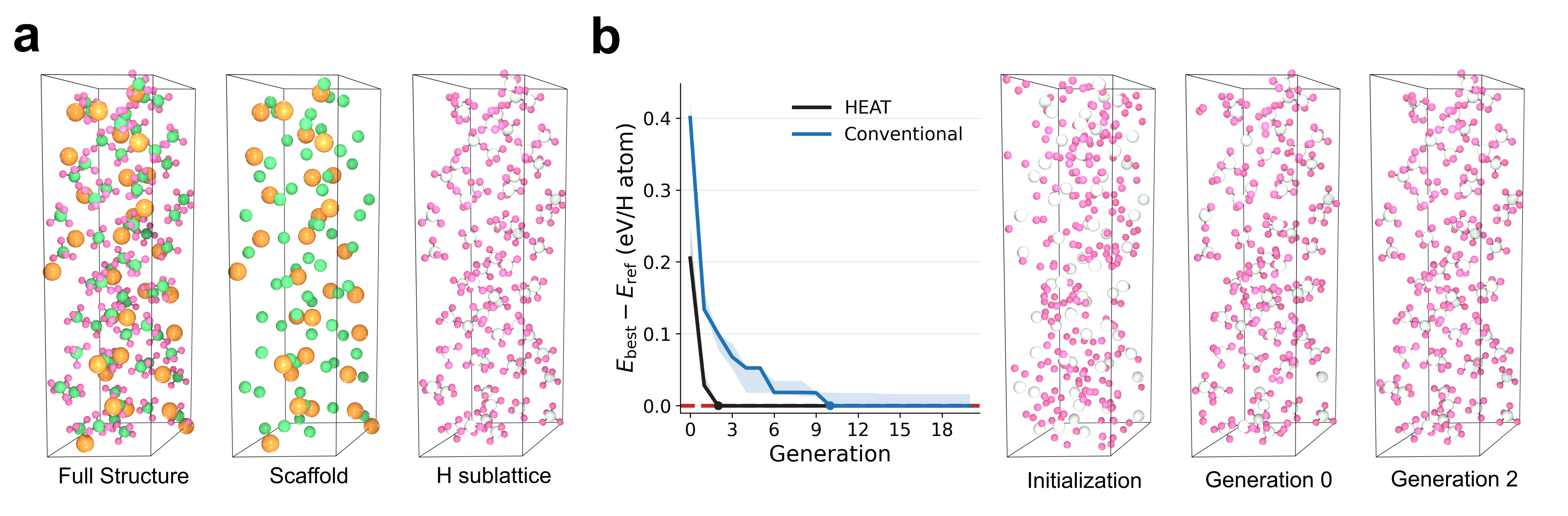}
    \caption{\textbf{Reconstruction of the H sublattice in $\mathbf{Mg(BH_4)_2}$ and comparison with a conventional GA.}
    \textbf{a} FAIRChem-relaxed reference structure and its decomposition into the supplied Mg--B scaffold and withheld 240-atom H sublattice. Mg, B, and H are orange, green, and pink, respectively, and B--H bonds are gray. White B atoms are retained in the isolated-sublattice view to better visualize the three-dimensional H network.
    \textbf{b} Cumulative best energy difference per H atom, $E_{\mathrm{best}}-E_{\mathrm{ref}}$, for HEAT (black) and the conventional GA using real atoms (blue) over the first 20 generations. Here, $E_{\mathrm{ref}}$ is the reference energy calculated by FAIRChem, normalized by the number of H atoms. The horizontal red dashed line marks zero energy difference. Solid curves show the run that first reaches the reference energy, and shaded bands span the minimum and maximum across five independent runs per method. Filled points mark the first generation that reaches the reference energy: generation 2 for HEAT and generation 10 for the conventional GA. The structures show a representative discrete random initialization before EAT relaxation and the best candidates at generations 0 and 2 from one of the HEAT runs. H atoms are pink, and B atoms shown for visual aid are white.}
    \label{fig:H_sublatt}
\end{figure*}

We first consider reconstructing a missing light-atom sublattice within a known heavier-atom framework. This setting is particularly favorable for a conventional genetic algorithm (GA): the fixed framework and known composition reduce the search largely to placing light atoms within a common host, making crossover a natural way to combine favorable arrangements from different parents. As shown below, even in this nearly ideal case for GA and using the same initialization procedure, population size, and final positional-relaxation criterion for both methods, HEAT outperforms the conventional GA.

This reconstruction problem is particularly relevant for hydrogen-containing crystals: because hydrogen has only one electron, its contribution to conventional X-ray scattering is weak, especially in the presence of heavier elements, and neutron diffraction is therefore often used to determine hydrogen positions. High-quality X-ray data combined with modern refinement methods can nevertheless recover accurate hydrogen coordinates in favorable cases~\cite{woinska2016hydrogen,haberl2021neutron}, while experimentally reported or database-derived structures may still contain well-resolved heavy-atom frameworks but missing, uncertain, or approximately assigned hydrogen positions~\cite{reents2026hydrogen}. Missing-hydrogen completion has recently also been studied using score-based diffusion models that inpaint H positions while holding the host structure fixed~\cite{reents2026hydrogen}.

Specifically, we reconstruct the 240-atom H sublattice of $\mathrm{Mg(BH_4)_2}$, a structurally complex light-metal borohydride of interest for hydrogen storage~\cite{filinchuk2009MgB2H8,zavorotynska2016magnesium}. The experimentally reported Mg--B--H structure~\cite{filinchuk2009MgB2H8} is first relaxed with FAIRChem and used as the reference. All H atoms are then removed, as their positions are often difficult to determine experimentally. The algorithms receive the relaxed Mg--B scaffold, unit cell, chemical formula, and hence the stoichiometric H count, but no reference H coordinates or bonding information (see Fig.~\ref{fig:H_sublatt}\textbf{a}).

This comparison deliberately gives HEAT and the conventional GA the same random starting distribution. Every generation-0 member and later random immigrant contains 240 fully occupied H atoms sampled from a $16\times16\times48$ feasible grid with a $1.0$~\AA{} minimum separation from the scaffold and from other initialized H atoms. HEAT then relaxes the positions and H--null mixture weights of the effective atoms, while the conventional GA only relaxes the positions. Both populations contain 30 members with 6 elites, 3 random immigrants, and 21 crossover children after generation zero, and both use tournament size 3. 

Figure~\ref{fig:H_sublatt}\textbf{b} compares five independent runs per method through generation 20. HEAT reaches the reference-energy criterion at generation 2 in all five runs. The conventional GA reaches it in three runs, at generations 10, 12, and 13; the remaining two do not reach it by generation 20. The discrete initialization is highly disordered, but relaxation in EAT's larger, continuous embedding space already produces recognizable local B--H coordination in generation zero.

For the highlighted runs, HEAT's faster convergence also translates into fewer interatomic-model evaluations. HEAT reaches the reference-energy criterion at generation 2 after 24,112 evaluations, compared with 65,431 evaluations at generation 10 for the conventional GA. Thus, in this comparison, the additional effective-atom relaxation reduces the total number of evaluations required to reach the reference criterion.

\subsection{Edge-dislocation-core prediction}
\label{sec:results:disloc}

\begin{figure*}[t]
    \centering
    \includegraphics[width=0.99\linewidth]{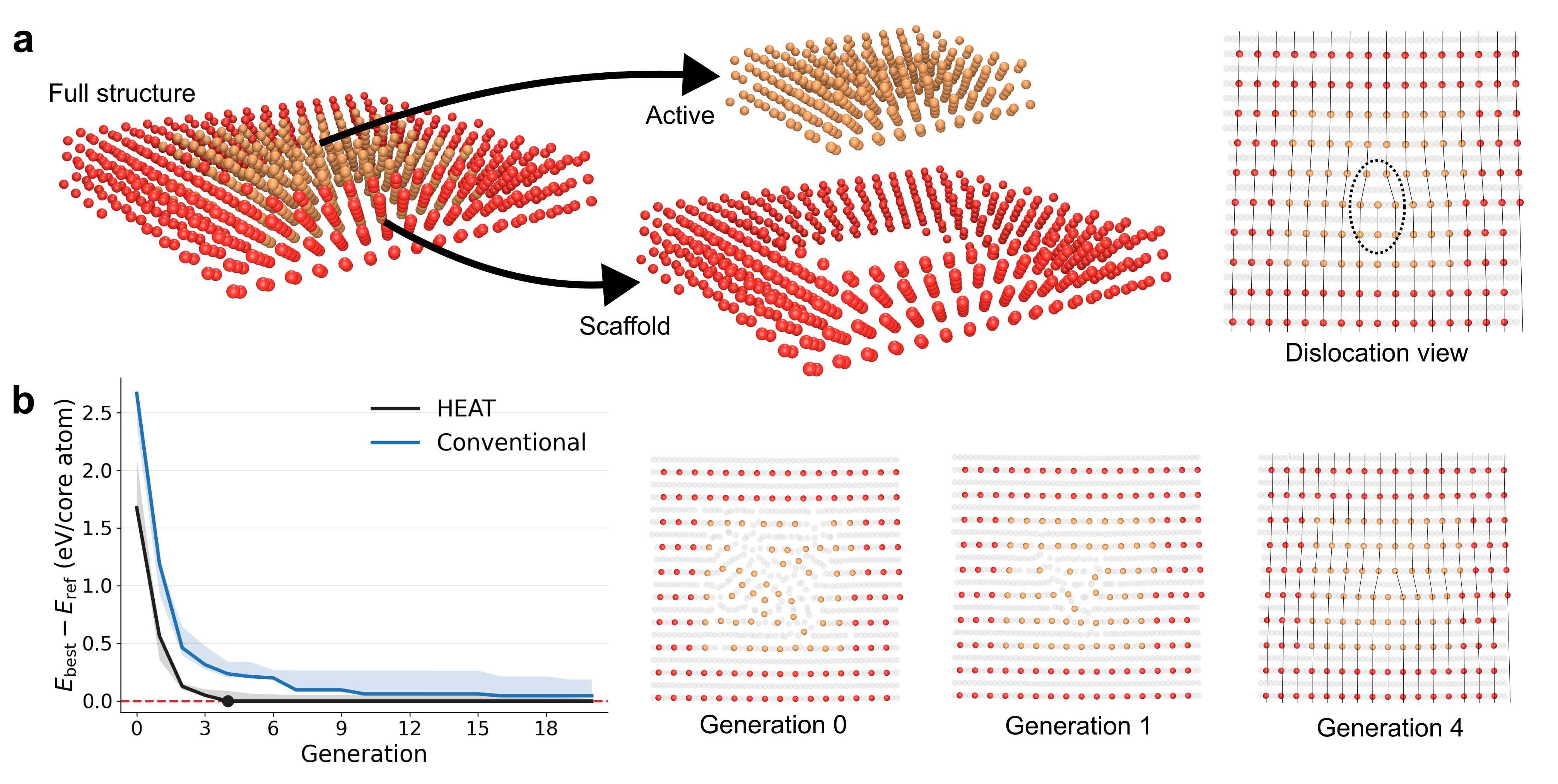}
    \caption{\textbf{Reconstruction of an edge-dislocation core in Fe and comparison with a conventional GA.}
    \textbf{a} Reference structure and its decomposition into the active region (bronze) and fixed scaffold (red). Starting from a configuration obtained from the NIST Interatomic Potentials Repository~\cite{becker2013nist}, the centered $25~\text{\AA}\times25~\text{\AA}$ active region is relaxed with FAIRChem while the scaffold and supercell lattice vectors are held fixed. The active region contains 364 atoms, and the rightmost view displays the extra half-plane. The dislocation core is encircled and contains roughly 48 atoms when accounting for deeper layers along the dislocation axis.
    \textbf{b} Cumulative best energy differences per core atom over the first 20 generations. Solid curves show the HEAT run (black) that first reaches the reference energy and the conventional-GA run (blue) with the lowest final energy, since none reaches the reference energy. Shaded bands span the minimum and maximum across five independent runs per method. The red dashed line marks zero energy difference, and the filled point marks HEAT reaching the reference energy at generation 4. The structures displayed are from the best HEAT run at generations 0, 1, and 4.}
    \label{fig:disloc}
\end{figure*}

We next consider a more challenging reconstruction problem: predicting an edge-dislocation core within a known surrounding crystal. Unlike the H-sublattice problem, neither the atomic arrangement within the core nor the number of atoms needed to form it is specified. Crossover must therefore reconcile competing arrangements and atomic populations, rather than simply combine placements within a common host framework.

Dislocation cores govern elementary processes such as glide and kink formation and therefore strongly influence the plastic response of crystalline materials~\cite{rodney2017dislocation}. The Burgers vector specifies the topological constraint imposed by a dislocation, while continuum elasticity predicts its far-field displacement. These constraints establish the surrounding deformation but do not determine the atomic structure of the compact, highly distorted core, where continuum elasticity breaks down. Reconstructing this core from the known surrounding crystal is therefore an important problem in materials theory.

Specifically, we consider an edge dislocation in Fe, starting from a configuration obtained from the NIST Interatomic Potentials Repository~\cite{becker2013nist}. This configuration was constructed using anisotropic elasticity and relaxed by low-temperature molecular dynamics followed by energy minimization using the Fe interatomic potential of Mendelev et al.~\cite{mendelev2003iron}. As in the H-sublattice demonstration, we further relax the database configuration with FAIRChem until the maximum force falls below $0.05$~eV/\AA{}, while holding the surrounding scaffold and cell fixed. The resulting structure and energy define the reference against which both HEAT and the conventional GA are evaluated.

Figure~\ref{fig:disloc}\textbf{a} shows the edge-dislocation reference structure and its decomposition into an active core and a fixed scaffold. The extracted section measures approximately $40\times40\times7$~\AA{}, with a centered $25\times25\times7$~\AA{} active core. We add 8~\AA{} of vacuum padding on each transverse side of the scaffold to separate periodic material images. Removing the 364 active core atoms leaves 605 fixed scaffold atoms. Neither HEAT nor the conventional GA receives the withheld core coordinates or atom count.

Random members in HEAT and the conventional GA are generated using the same initialization procedure. The proposed atom count is drawn from a Poisson distribution whose mean is the active-region volume times the number density inferred from the supplied scaffold, without using the reference count, and clipped between one and twice this estimate. Proposed atoms are distributed as evenly as possible among a $6\times6\times6$ grid, with random positions within each grid cell. Sites are then pruned to enforce a minimum separation of $1.0$~\AA{} from other proposed atoms and the scaffold, which can reduce the initial atom count. Both populations contain 30 members, with 3 elites, 8 random immigrants, and 19 crossover children per subsequent generation, using tournament size 2. Each conventional-GA crossover child has a probability of 0.30 of undergoing one mutation: rattle, relocation, insertion, or deletion. Before scoring, every newly generated candidate undergoes the same final positional relaxation with the scaffold and cell fixed; elites retain their previously relaxed scores.

The results distinguish the methods across independent runs (Fig.~\ref{fig:disloc}\textbf{b}). HEAT reaches the reference-energy criterion in two of five runs, at generations 4 and 15, whereas none of the five conventional-GA runs reaches it by generation 20. The highlighted HEAT run reaches the criterion after 50,629 model evaluations. By generation 20, the highlighted conventional run has used 156,634 evaluations but remains 0.045~eV per core atom (0.36~eV per layer) above the reference. In the highlighted HEAT run, the disordered generation 0 core develops ordered rows by generation 1 and recovers the full edge-dislocation core by generation 4. This progression illustrates how HEAT can complete a dislocation structure whose surrounding framework is established from far-field elasticity, providing a rapid method for determining the atomic structure of the unknown core.

Allowing atomic positions and populations to adjust together during relaxation is central to this rapid recovery. The conventional GA can change the atom count through crossover and insertion or deletion mutations, but its subsequent local relaxation proceeds at fixed population. In contrast, continuous transfer of weight between Fe and null in HEAT allows atoms to appear or disappear as the structure relaxes, opening additional pathways for changes in bonding topology and for reconciling competing atomic arrangements inherited from the parents. In the highlighted runs, HEAT thereby recovers the reference core with substantially fewer model evaluations than the conventional GA expends without reaching it.

\subsection{Structure Prediction from Ptychography Information}
\label{sec:ptycho}

\begin{figure*}[t]
    \centering
    \includegraphics[width=0.99\linewidth]{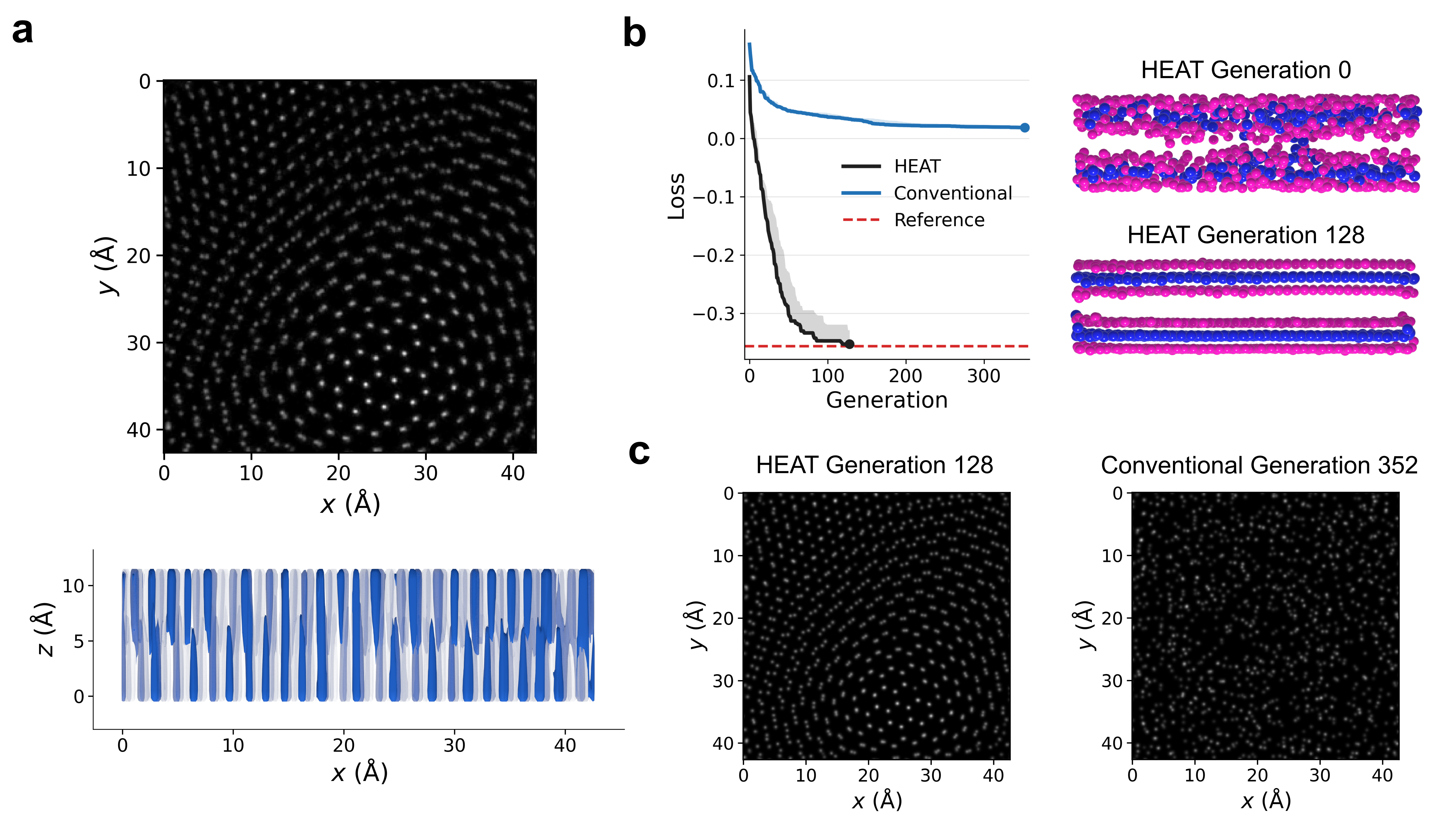}
    \caption{\textbf{Structure prediction of twisted-bilayer $\mathbf{WSe_2}$ from experimental ptychography data and comparison with a conventional GA.}
    \textbf{a} Experimental phase volume from~\cite{nguyen2024wse2}, summed over the 12 measured depth slices (top), and an orthographic depth view of its normalized-intensity isosurface at contour value $I^{\mathrm{exp}}=0.20$ (bottom). The elongated contours illustrate the weak depth localization. Opacity decreases with distance along the viewing direction and the surface is clipped at the measured depth boundaries.
    \textbf{b} Cumulative best loss for three independent runs per method, shown through generation 128 for HEAT (black) and generation 352 for the conventional GA (blue), at approximately equal energy-evaluation budgets. Solid curves show the single run with the lowest loss at each method's comparison endpoint; shaded bands span the minimum and maximum across its three runs. Filled circles mark these comparison endpoints. The loss combines formation energy per atom with image, composition, and sample-box penalties (Eq.~\eqref{eq:wse2_total_objective}, $\beta=0$, see Methods). The red dashed line marks the loss of an independently constructed ideal reference $\mathrm{WSe_2}$ bilayer fitted to the experimental phase volume and relaxed under the same objective. Depth views show the best candidates at generations 0 and 128 from a representative HEAT run. W and Se are blue and magenta, respectively.
    \textbf{c} Simulated phase volumes (compare with \textbf{a}), summed over the same 12 slices, for the best candidates at generations 128 and 352 from representative HEAT and conventional-GA runs, respectively. Experimental and simulated projections use a common intensity scale. All spatial views show the central $42.72\times42.72$~\AA{} region of the experimental sample.}
    \label{fig:ptycho}
\end{figure*}

As our final application, we apply HEAT to three-dimensional atomic-structure reconstruction from electron-ptychography data. Electron ptychography can achieve deep-sub-\AA{}ngstr\"om lateral resolution and picometer-scale precision~\cite{jiang2018ptychography,chen2020mixedstate,chen2021ptychography}, while multislice methods additionally encode depth information. However, the depth resolution remains substantially poorer, leaving three-dimensional atomic coordinates incompletely constrained~\cite{gao2017threeD,chen2024interstitial,dong2025depth,karapetyan2026metrology}. Related approaches have combined incomplete microscopy data with energetic modeling and Bayesian genetic algorithms~\cite{debacker2022bayesian}, and also tracked the three-dimensional evolution of Pt nanoparticle facets during catalysis by constructing crystalline starting structures from projected atom counts and relaxing them with molecular dynamics~\cite{jenkinson2026operando}. Recent work retrieved three-dimensional coordinates in twisted-bilayer $\mathrm{WSe_2}$ using sample tilt to separate overlapping atoms, structural constraints, and manual estimates for ambiguous sites~\cite{huang2026wse2}. HEAT instead combines the ptychography data with FAIRChem energies to optimize positions, species, and atom count, without requiring a crystalline starting structure, tilting the sample to aid peak finding, or prescribing atomic layers and bonds.

We focus on reconstructing a $42.72\times42.72$~\AA{} interior region of the twisted-bilayer $\mathrm{WSe_2}$ sample from~\cite{nguyen2024wse2, lee2025ptyrad}. The experimental input is a three-dimensional phase volume, which constrains the plausible locations of atoms. Figure~\ref{fig:ptycho}\textbf{a} shows the corresponding depth-summed image and a depth view of the measured-field isosurface. The phase field has much better resolution in the $x$--$y$ plane than along $z$, tightly constraining lateral atomic positions while leaving substantial uncertainty in their depths. Recovering a discrete structure therefore requires resolving both this depth ambiguity and the unknown number and identities of atoms across a region containing more than 1000 atoms.

The optimization covers a larger $48.82\times48.82$~\AA{} region of the sample, surrounded by approximately 8~\AA{} of vacuum padding to separate periodic images. This padding introduces artificial edges that are absent from the experimental sample and can distort the optimized structure. We therefore optimize a region larger than the area of interest, keeping the central $42.72\times42.72$~\AA{} region away from these edges to reduce their influence. This central crop is used for visualization and the projection comparison below, while all reported optimization losses are evaluated over the full measured region.

We formulate the reconstruction as a Bayesian inference problem, combining the experimental imaging evidence with chemical information supplied by the MLIP. For an atomic configuration $X$ and measured phase volume $I$, we write the posterior probability as $P(X| I,M)\propto P(I| X)P(X| M)$, where $M$ denotes the MLIP information. The image likelihood $P(I| X)$ measures agreement with the experimental data, while the prior $P(X| M)$ favors energetically plausible structures. Maximizing this posterior is equivalent to minimizing the sum of the corresponding negative logarithms, providing a common objective for balancing imaging evidence and chemical plausibility. Finally, we introduce additional soft priors to favor the global W:Se ratio of $1:2$ and to confine atoms to the measured sample region, without imposing an exact atom count or exact stoichiometry.

To construct the energy-based prior, we evaluate the formation energy per atom, $E_{\mathrm M}(X)$, using FAIRChem and elemental reference energies. Maximizing the entropy over configurations subject to a constraint on the mean energy above the MLIP global minimum gives $P(X| M)\propto\exp[-\alpha E_{\mathrm M}(X)]$, where $\alpha>0$ is an inverse energy scale controlling the preference for low-energy structures. This scale characterizes our energetic prior and need not correspond to a physical temperature. Combining this prior with the image likelihood and dividing the negative log posterior by $\alpha$ gives the loss to be minimized as $\mathcal{L} = E_{\mathrm M}(X)-\alpha^{-1}\ln P(I| X)$, supplemented by the composition and confinement penalties introduced above.

To model the image likelihood, we first predict the phase volume associated with a candidate atomic configuration. Evaluating a full electron-scattering simulation and ptychographic reconstruction for every candidate would be computationally expensive. We instead use a differentiable approximation based on a precomputed library of isolated-atom projected potentials created using PtyRAD~\cite{lee2025ptyrad}. Each elemental lateral profile is fitted with a sum of two Gaussians, representing its narrow central peak and broader tails, and combined with a Gaussian depth response of standard deviation $4$~\AA{}. Summing these contributions at the candidate atomic positions gives the predicted phase volume. Given this model phase volume, we next treat the voxelwise residuals as independent, with a probability density proportional to the exponential of a negative Huber penalty. The resulting residual distribution has a Gaussian center and exponential tails, accommodating small discrepancies as approximately Gaussian errors while reducing the influence of larger deviations, including those arising from imperfections in the forward model. The negative log likelihood $-\ln P(I| X)$ is therefore proportional to the summed Huber penalties, up to a configuration-independent constant. The normalization, penalty weights, and remaining implementation details are given in Methods.

To perform the structural inference, both HEAT and the conventional GA minimize the same loss objective, with HEAT employing effective-atom mixtures of W, Se, and null during relaxation. Both methods use the experimental phase field to guide initialization, placing candidate atoms in high-intensity regions. During initialization, HEAT first performs an image-fitting stage that adjusts positions and continuous species weights, then includes the interatomic energy during EAT relaxation, whereas the conventional GA assigns discrete W and Se atoms to locations sampled from the same candidate region and locally relaxes only their positions.

In later generations, as described above, HEAT uses whole-structure superposition followed by EAT relaxation, while the conventional GA uses cut-and-splice crossover and discrete mutations to change the atomic population and arrangement. Both algorithms use populations of 50, with 5 elites, 13 random immigrants, and 32 crossover children per generation, and tournament sizes of 4 and 2 for selecting the first and second parents, respectively. For HEAT, selection and all reported losses are based on the final real-atom structures, excluding the entropy term used during syntropization (Eq.~\eqref{eq:wse2_total_objective} with $\beta=0$ in Methods).

To establish a reference, we constructed two ideal $\mathrm{WSe_2}$ monolayers from a Materials Project bulk structure, then fitted their orientations, translations, heights, and small in-plane strains to the experimental phase volume. We subsequently relaxed all atomic positions under the same combined energy, image, composition, and confinement loss used for the GA candidates, holding the species, atom count, and cell fixed, to obtain a reference loss of $-0.3567$. 

Figure~\ref{fig:ptycho}\textbf{b} compares three independent runs per method through generation 128 for HEAT and generation 352 for the conventional GA. At these endpoints, all three HEAT runs achieve lower losses ($-0.353$ to $-0.329$) than all three conventional-GA runs ($0.019$ to $0.023$). The highlighted HEAT run improves from $0.11$ to $-0.353$, first attained at generation 119, whereas the highlighted conventional run reaches $0.019$. Their formation energies are $-0.391$ and $-0.089$~eV/atom, and their weighted image penalties are $0.035$ and $0.103$, respectively. HEAT remains close to the independently constructed reference values of $-0.387$~eV/atom and $0.027$.

The comparison endpoints correspond to approximately equal energy evaluation budgets: HEAT uses $529{,}000$--$544{,}000$ energy evaluations through generation 128, compared with $544{,}000$--$546{,}000$ for the conventional GA through generation 352. 

The depth views to the right of the convergence plot show the best structures at generations zero and 128 from the representative HEAT run, with W in blue and Se in magenta. The generation-zero candidate already forms two broad layers, but their internal height ordering is disordered. By generation 128, the retained best structure resolves the six atomic planes of two twisted Se--W--Se trilayers. The full structure contains 1497 atoms; the displayed interior crop contains 1150 atoms, comprising 389 W and 761 Se.

Figure~\ref{fig:ptycho}\textbf{c} compares the depth-summed phase volumes for the $42.72\times42.72$~\AA{} region predicted by the approximate ptychography reconstruction for the best candidates at generations 128 and 352 from the representative HEAT and conventional-GA runs, respectively. HEAT closely reproduces the measured projected pattern, with a Pearson correlation of $0.934$, compared with $0.125$ for the conventional GA. Residual intensity mismatches remain at only a few locations, indicating potential W or Se vacancies. A simple post-processing step may suffice to correct such point defects. Together with the recovered trilayer structure and lower objective, the results show that HEAT accesses physically favorable, image-consistent configurations that the conventional GA fails to find within the comparison window.

\section{Discussion}

HEAT combines genetic search with continuous optimization of atomic positions, identities, and occupancies. Its superposition crossover allows atoms from both parents to contribute to a child, while redundant or unfavorable atoms disappear when doing so lowers the loss during relaxation. Across five independent runs per method, HEAT reaches the reference-energy criterion in all five H-sublattice runs and two edge-dislocation runs, compared with three and zero runs, respectively, for the conventional GA within 20 generations. The fastest HEAT runs also reach the criterion in fewer generations. The largest application, twisted-bilayer $\mathrm{WSe_2}$, shows a substantial improvement in both formation energy and agreement with the experimental ptychographic data. These results demonstrate the value and efficiency of the effective-atom representation for reconstructing missing sublattices, determining dislocation-core structures---with potential extensions to grain boundaries and interfaces---and recovering atomic structures from ptychographic data.

HEAT introduces computational overhead relative to conventional genetic algorithms by optimizing chemical identities and occupancies alongside atomic positions, combining both parents' atoms during crossover, and following effective-atom optimization with a final real-atom relaxation. However, across all three applications, this overhead is more than offset by HEAT's greater effectiveness in exploring chemical orderings and bonding topologies. The additional operations do not necessarily increase the net cost per generation, because continuous changes in atomic identity and presence can provide more efficient relaxation pathways. In the highlighted H-sublattice and edge-dislocation runs, HEAT reaches the reference-energy criterion with fewer model evaluations than the conventional GA uses in the comparisons shown. For the ptychographic reconstruction, HEAT requires approximately 2.6 times as many energy evaluations per generation, but achieves significantly better structural quality than the conventional GA at approximately equal energy evaluation budgets. Estimated runtimes on a single NVIDIA RTX 5090 are approximately 1~h and 3.9~h for reference recovery in the highlighted H-sublattice and edge-dislocation HEAT runs, respectively, and 112~h for the representative $\mathrm{WSe_2}$ HEAT run through generation 128. We find the greatest benefit in our largest system. For smaller problems that a conventional GA already solves readily, HEAT may offer relatively little advantage. 

Looking forward, several natural extensions could broaden HEAT's applicability. Allowing the lattice vectors to vary would extend the present fixed-cell searches to simultaneous optimization of atomic structure and supercell geometry. For systems without a supplied scaffold or experimental field to establish a common coordinate frame, exploring relative translations and rotations of the parents before superposition could enable crossover to identify and combine additional compatible structural motifs. The effective-atom representation accommodates both extensions, providing a foundation for their development within HEAT. We also plan to explore extending HEAT to ptychographic reconstruction of amorphous systems containing more than $100{,}000$ atoms, thereby bringing atomically resolved structural inference to complex disordered material systems of direct technological relevance.

\section{Methods}
\label{sec:methods}

\subsection{Effective-atom representation and HEAT}

For effective atom $I$, let $x_{I\alpha}\geq0$ be its weight for real species $\alpha$. The real-species weights satisfy $\sum_\alpha x_{I\alpha}\leq1$, and the remaining weight
\begin{equation}
    x_{I\emptyset}=1-\sum_\alpha x_{I\alpha}
\end{equation}
is assigned to null~\cite{tahmassebpur2025eat}. The limits $x_{I\emptyset}=1$ and $x_{I\alpha}=1$ represent an absent atom and a conventional atom of species $\alpha$, respectively. The interatomic energy and application-specific penalties are differentiated with respect to both $x_{I\alpha}$ and the position $\mathbf r_I$, allowing identity, presence, and geometry to relax together.

Syntropization adds the Tsallis penalty~\cite{tsallis1988possible,tahmassebpur2025eat}
\begin{equation}
    S_q=\sum_I
    \frac{1-\sum_\alpha x_{I\alpha}^{q}-x_{I\emptyset}^{q}}{q-1}
    \label{eq:tsallis_syntropization}
\end{equation}
to the physical objective with coefficient $\beta$. Increasing $\beta$ favors vertices of the species--null simplex. After syntropization, any remaining mixed-species effective atoms are assigned to the species with the largest weight; effective atoms assigned to null are removed. Thus every structure admitted to the genetic population is a conventional discrete structure, and fractional occupancies never enter the final comparison score.

Generation zero contains $P$ independently initialized and optimized random members. Each subsequent generation retains the $E$ lowest-score members without reevaluation (elites), introduces $R$ newly initialized random immigrants, and creates $C=P-E-R$ crossover children. The same initialization setting controls generation zero and all later immigrants. Elitism preserves the best structures, while immigrants provide continuing exploration~\cite{greenstein2023bestpractices}.

Parents are chosen only by fitness through two tournament selections~\cite{miller1995tournament}. For independently sampled tournaments $T_A$ and $T_B$,
\begin{align}
    A &= \arg\min_{X\in T_A}\mathcal L(X),\\
    B &= \arg\min_{X\in T_B}\mathcal L(X),
\end{align}
where $\mathcal L(X)$ is the loss function (equivalently the negative fitness). We clarify that parents are discrete structures so the entropy term in the loss associated with syntropization is zero. The winner $A$ is excluded from the eligible population when $T_B$ is drawn, ensuring distinct parents; other competitors may occur in both tournaments.

To cross parents $A$ and $B$, each optimizable atom is converted into an effective atom with weight $\kappa$ on its original element and weight $1-\kappa$ on null:
\begin{align}
    x'_{I\alpha} &= \kappa\,\delta_{\alpha,s_I},
    &&\alpha\neq\emptyset,\\
    x'_{I\emptyset} &= 1-\kappa,
    \label{eq:occupancy_scaling}
\end{align}
where $s_I$ is the original element of atom $I$ and $0<\kappa<1$. The two complete sets of optimizable effective atoms are concatenated in their common scaffold and cell. Effective atoms that overlap are reduced to one, small random perturbations are applied to occupancy and position, and zero-occupancy reservoir effective atoms are added when needed. The resulting child is optimized with EAT, hard-rounded, and relaxed as a conventional atomic structure. We use $\kappa=0.5$, so the child initially retains half of the real occupancy carried by every site in each parent. 

After hard rounding, null sites are removed and the positions of the remaining active atoms are relaxed. Copied elites retain their previously computed scores without further relaxation.

\subsection{Conventional genetic baseline}
\label{sec:methods:conventional}

The baseline is a conventional fixed-cell memetic GA assembled from conventional selection, cut-and-splice, mutation, and local-relaxation operators, with any supplied scaffold held fixed. Its candidates contain only fully occupied hard atoms. In the H-sublattice and edge-dislocation benchmarks, it uses the same hard random-member initialization scheme as HEAT. The H benchmark initializes the known stoichiometric number of atoms; the dislocation benchmark draws a Poisson population estimate from the supplied scaffold density. Initial and repaired atoms obey the same minimum-distance cutoff.

Parent selection uses the same best-of-tournament rule. In the H-sublattice and edge-dislocation benchmarks, crossover is conventional periodic real-space cut-and-splice: a random Cartesian plane through a random fractional-cell origin is drawn, the child takes one periodic half-space from parent $A$ and the opposite half-space from parent $B$, and the resulting positions are repaired for overlaps and allowed atom-count limits. Each crossover child is mutated with probability 0.30. H mutations either rattle a subset of atoms or relocate one atom; dislocation mutations may additionally insert or delete an atom. No mutation is applied to elites or random immigrants. The microscopy-frame crossover and initialization used for the $\mathrm{WSe_2}$ baseline are specified below in subsection \emph{Conventional GA baseline for ptychographic reconstruction}.

Every non-elite baseline member is locally relaxed and scored under the same loss function used for HEAT. The methods differ in their local search before hard scoring and in their crossover and mutation operators. In the H-sublattice and edge-dislocation benchmarks, both methods use a population of 30 and the same numbers of elites, random immigrants, and crossover children. The initialization and population choices for each application are given below. For the H-sublattice and edge-dislocation benchmarks, we compare five independent runs per method over generations 0--20. A run meets the reference-energy criterion when its best formation energy per total atom satisfies $E_f-E_f^{\mathrm{ref}}<10^{-4}$~eV/atom, including values below the reference. This criterion uses all atoms, including the scaffold, whereas the figures normalize energy differences per H atom or per core atom. It measures energy agreement rather than structural identity; the displayed structures illustrate recovery in the highlighted HEAT runs.

\subsection{Hydrogen-sublattice reconstruction protocol}
\label{sec:methods:sublatt}

The complete experimentally reported $\mathrm{Mg(BH_4)_2}$ structure of~\cite{filinchuk2009MgB2H8} is first relaxed with FAIRChem, including cell relaxation, and used as the reference. All 240 H atoms are then removed, leaving the Mg--B framework fixed for candidate calculations. Each effective atom in the unknown sublattice spans H and null, and the continuous effective number of H atoms is
\begin{equation}
    N_{\mathrm{H}}
    =
    \sum_I x_{I\mathrm{H}}.
\end{equation}

The loss minimized during each HEAT relaxation is
\begin{equation}
    \mathcal{L}
    =
    E_f
    +
    \lambda P_{\mathrm{comp}}
    +
    \beta S_q,
    \label{eq:loss_H_sublatt}
\end{equation}
where $E_f$ is the formation energy per atom and $P_{\mathrm{comp}}$ enforces the target stoichiometry. The formation energy is
\begin{equation}
    E_f
    =
    \frac{
        E
        -\mu_{\mathrm{Mg}}N_{\mathrm{Mg}}
        -\mu_{\mathrm{B}}N_{\mathrm{B}}
        -\mu_{\mathrm{H}}N_{\mathrm{H}}
    }{
        N_{\mathrm{Mg}}+N_{\mathrm{B}}+N_{\mathrm{H}}
    },
    \label{eq:H_sublatt_formation_energy}
\end{equation}
where $N_{\mathrm{Mg}}$ and $N_{\mathrm{B}}$ are the numbers of Mg and B atoms in the fixed scaffold, respectively.

The composition penalty is
\begin{equation}
    P_{\mathrm{comp}}
    =
    \sum_{\alpha\in\{\mathrm{Mg,B,H}\}}
    \left(
        x_\alpha-x_\alpha^{\mathrm{target}}
    \right)^2,
\end{equation}
where
\begin{equation}
    x_\alpha
    =
    \frac{N_\alpha}
    {N_{\mathrm{Mg}}+N_{\mathrm{B}}+N_{\mathrm{H}}},
\end{equation}
and
\begin{equation}
    x_{\mathrm{Mg}}^{\mathrm{target}}=\frac{1}{11},
    \qquad
    x_{\mathrm{B}}^{\mathrm{target}}=\frac{2}{11},
    \qquad
    x_{\mathrm{H}}^{\mathrm{target}}=\frac{8}{11}.
\end{equation}
The composition term fixes only the global stoichiometry and provides no information about H coordinates or bonding. Its coefficient is $\lambda=5$. Random members begin with exactly 240 one-hot H atoms chosen without replacement from feasible locations on a $16\times16\times48$ fractional grid. Feasible locations are at least 1.0~\AA{} from the scaffold, and the selected H atoms are also separated by at least 1.0~\AA{}. This generator is used identically for every generation-0 member and every later random immigrant in HEAT and the conventional baseline. HEAT may subsequently change the realized population through H--null relaxation; the composition penalty then softly restores the known global stoichiometry.

After the continuous stage, the candidate is hard-rounded and locally relaxed with the scaffold and cell fixed; the conventional baseline uses the same final relaxation and force criterion. Ranking uses the hard formation energy plus the hard composition term, with no entropy contribution. Both searches use $P=30$, with 6 elites, 3 random immigrants, 21 crossover children, and tournament size 3.

\subsection{Edge-dislocation-core reconstruction protocol}
\label{sec:methods:disloc}

The edge-dislocation reference is initialized as a configuration from~\cite{becker2013nist}. Atoms in a centered $25~\text{\AA}\times25~\text{\AA}$ active prism are relaxed with FAIRChem while the surrounding scaffold and cell are fixed. The extracted material section spans approximately $40~\text{\AA}\times40~\text{\AA}$ transversely and is periodic along the dislocation line, with a repeat of approximately 7~\AA{}. We add 8~\AA{} of vacuum on each transverse side to separate periodic images. The system contains 605 fixed scaffold atoms and 364 relaxed active atoms. The active atoms are removed before either genetic search begins.

For both algorithms, a random member count $N_0$ is drawn as
\begin{equation}
    N_0\sim\operatorname{Poisson}(\rho_{\mathrm{scaff}}V_{\mathrm{act}}),
\end{equation}
clipped to at least one atom and at most twice the density estimate. Positions are stratified across a $6\times6\times6$ partition of the active prism and repaired using a 1.0~\AA{} minimum distance from other candidate atoms and the scaffold. This initialization is used for every generation-0 member and random immigrant and does not use the withheld reference population.

Each effective atom in the active region spans Fe and null. A random effective-atom initialization can collapse to the all-null configuration before an ordered core forms, so we introduce a soft lower bound on the total non-null occupancy. The objective is
\begin{equation}
    \mathcal{L}
    =
    E_f
    +
    \lambda P_{\mathrm{occ}}
    +
    \beta S_q,
    \label{eq:loss_disloc}
\end{equation}
where $P_{\mathrm{occ}}$ penalizes insufficient occupancy. The continuous effective number of active atoms is
\begin{equation}
    N_{\mathrm{act}}
    =
    \sum_I x_{I\mathrm{Fe}},
    \label{eq:disloc_active_population}
\end{equation}
and the formation energy is
\begin{equation}
    E_f
    =
    \frac{
        E-\mu_{\mathrm{Fe}}
        \left(N_{\mathrm{scaff}}+N_{\mathrm{act}}\right)
    }{
        N_{\mathrm{scaff}}+N_{\mathrm{act}}
    },
    \label{eq:disloc_formation_energy}
\end{equation}
where $N_{\mathrm{scaff}}$ is the number of fixed scaffold atoms.

The occupancy penalty is
\begin{equation}
    P_{\mathrm{occ}}
    =
    \operatorname{softplus}_{k}
    \left(
        \gamma\rho_{\mathrm{scaff}}V_{\mathrm{act}}
        -
        N_{\mathrm{act}}
    \right),
    \label{eq:occupancy_penalty}
\end{equation}
where
\begin{equation}
    \rho_{\mathrm{scaff}}
    =
    \frac{N_{\mathrm{scaff}}}{V_{\mathrm{scaff}}}.
\end{equation}
Here, $V_{\mathrm{scaff}}$ and $V_{\mathrm{act}}$ are the volumes of the supplied material outside and inside the active prism, $k$ controls the softplus sharpness, and $\gamma\in(0,1]$ sets the minimum encouraged occupancy. The density estimate uses only the scaffold, not the reference count. We use $k=0.5$ and $(\lambda,\gamma)=(20,0.3)$. These deliberately loose bounds prevent early collapse to an empty active region without fixing the final population.

Null effective atoms are added after optimization when needed so that EAT can still spawn in atoms during later relaxation. Crossover halves the parent occupancies, superimposes and perturbs the parents, and prunes overlaps before EAT relaxation. The resulting hard Fe atoms are finally relaxed under the same fixed-scaffold protocol used for the reference and the conventional GA baseline. The hard formation energy plus occupancy-lower-bound term determines ranking. Both algorithms use $P=30$, with 3 elites, 8 immigrants, 19 crossover children, and tournament size 2.

\subsection{Ptychographic reconstruction protocol}
\label{sec:methods:ptycho}

The reconstruction protocol optimizes against the measured three-dimensional phase volume directly. The data remain on their native voxel grid; effective atoms are not assigned to individually fitted experimental blobs. The input is a centered $320\times320$ pixel crop containing all 12 measured depth slices, with sampling $(\Delta z,\Delta y,\Delta x)=(1.0,0.152564,0.152564)$~\AA{}. The measured region therefore spans $48.82\times48.82\times12$~\AA{}. Adding approximately 8~\AA{} of vacuum on each side, with lateral padding snapped to whole pixels, gives a fixed $64.69\times64.69\times28$~\AA{} periodic cell. Atomic centers are softly confined to the measured sample box. All measured voxels contribute to the image loss, while synthetic vacuum voxels are excluded. 

A single background and intensity scale are determined over the entire measured crop: subtract its first percentile, clip negative values to zero, and divide by the 99.5th percentile of the resulting field. The normalized target $I_v^{\mathrm{exp}}$ is retained in the linear domain, including values above one; individual slices are not normalized separately.

Each effective atom spans W, Se, and null. Let $o_I=1-x_{I\emptyset}$, $N_\alpha=\sum_I x_{I\alpha}$, and $N=N_{\mathrm{W}}+N_{\mathrm{Se}}$. At voxel $v$, the predicted image is
\begin{equation}
    \widehat I_v
    =g\sum_I\sum_{\alpha\in\{\mathrm{W,Se}\}}
    x_{I\alpha}
    G_\alpha^{xy}(\boldsymbol{\rho}_v-\boldsymbol{\rho}_I)
    R(z_v-z_I),
    \label{eq:wse2_forward_image}
\end{equation}
where $\boldsymbol{\rho}$ denotes lateral position. Each $G_\alpha^{xy}$ is a sum of core and tail Gaussians whose widths and relative integrated weights are taken from fits to isolated-atom projected-potential tables, with an additional 0.08~\AA{} lateral Gaussian blur. The shared depth response is
\begin{equation}
    R(\zeta)=\frac{\Delta z}{\sqrt{2\pi}\sigma_z}
    \exp\!\left[-\frac{\zeta^2}{2\sigma_z^2}\right],
    \qquad \sigma_z=4~\text{\AA}.
    \label{eq:wse2_depth_response}
\end{equation}
The lateral convolution is periodic; the depth response does not wrap across the cell. Its tails outside the measured depth window are not renormalized. A common factor $g$ sets the combined peak of one isolated W atom, centered on a lateral grid point and a depth plane, to one. 

The loss is
\begin{equation}
    \begin{split}
        \mathcal L={}&E_f+\lambda_{\mathrm{img}}P_{\mathrm{img}}
        +\eta P_{\mathrm{comp}}\\
        &+\lambda_{\mathrm{box}}P_{\mathrm{box}}+\beta S_q,
    \end{split}
    \label{eq:wse2_total_objective}
\end{equation}
with formation energy per atom
\begin{equation}
    E_f=\frac{E-\mu_{\mathrm W}N_{\mathrm W}
                    -\mu_{\mathrm{Se}}N_{\mathrm{Se}}}{N},
    \label{eq:wse2_formation_energy}
\end{equation}
where $E$ is the FAIRChem energy and $\mu_\alpha$ are elemental reference energies. The image penalty is the mean Huber loss over the measured voxels $\mathcal V$,
\begin{equation}
    P_{\mathrm{img}}=\frac{1}{|\mathcal V|}
    \sum_{v\in\mathcal V}\rho_\delta(\widehat I_v-I_v^{\mathrm{exp}}),
    \qquad \delta=0.20,
    \label{eq:wse2_image_loss}
\end{equation}
where $\rho_\delta(t)=t^2/2$ for $|t|\leq\delta$ and $\delta(|t|-\delta/2)$ otherwise. The global composition penalty is
\begin{equation}
    P_{\mathrm{comp}}=\sum_{\alpha\in\{\mathrm{W,Se}\}}
    \left(\frac{N_\alpha}{N}-f_\alpha^{\mathrm{target}}\right)^2,
    \label{eq:wse2_composition_penalty}
\end{equation}
with $f_{\mathrm W}^{\mathrm{target}}=1/3$ and $f_{\mathrm{Se}}^{\mathrm{target}}=2/3$. For sample-box bounds $\ell_d,u_d$ in Cartesian direction $d$,
\begin{equation}
    \begin{split}
        P_{\mathrm{box}}={}&\frac1N\sum_I o_I\sum_{d=x,y,z}
        \bigl[\operatorname{softplus}_k(\ell_d-r_{Id})^2\\
        &+\operatorname{softplus}_k(r_{Id}-u_d)^2\bigr],
    \end{split}
    \label{eq:wse2_box_penalty}
\end{equation}
with $\operatorname{softplus}_k(t)=k^{-1}\ln(1+e^{kt})$. The configured weights are $\lambda_{\mathrm{img}}=10$, $\eta=10$, and $\lambda_{\mathrm{box}}=50$, with $k=8$~\AA$^{-1}$. The composition term is a soft guide; neither exact stoichiometry nor a target atom count is enforced. 

For generation zero and random immigrants, an initialization mask is obtained by smoothing the square root of the normalized target and thresholding at its median. Candidate locations are selected in intensity order. Each selected location excludes further candidates within a cylinder centered on that location, with lateral radius 0.75~\AA{} and total height 6~\AA{} ($\pm3$~\AA{} along the depth direction). The selected locations are then perturbed by uniform jitter of $\pm0.10$~\AA{} per Cartesian coordinate, accepted only within the initialization region. The proposal count is set by a density of 0.20~\AA$^{-3}$ times the feasible measured volume, excluding added vacuum. Each site allows W, Se, or null. An initial 500-step image prefit jointly optimizes positions and soft species probabilities using Adam, the image, composition, and sample-box terms, a short-range repulsion, and a gradually increasing $q=2$ entropy penalty. This prefit omits the interatomic energy and passes all candidate effective atoms, still softly occupied, into the energy-inclusive EAT optimization.

Syntropization then uses projected quasi-Newton relaxation of Eq.~\eqref{eq:wse2_total_objective}, with $q=2$ and $\beta$ increasing from $10^{-3}$ by a factor of 1.5 until the mean entropy per site falls below $10^{-3}$. Each site is rounded to its largest W, Se, or null component; null atoms are removed, and positions are relaxed with species and cell fixed under the same physical objective. The final positional stage uses a projected-gradient tolerance of $3\times10^{-3}$, an iteration limit of 500, and a cost-plateau stopping criterion. Selection uses the final hard objective without the entropy term. The same image renderer is used in prefit, syntropization, and final relaxation.

Three independent runs are performed for each method. The population contains $P=50$ members. After generation zero, each generation retains 5 elites, adds 13 random immigrants, and creates 32 crossover children. Parent tournaments have sizes 4 and 2, excluding the first winner from the second tournament. Whole-structure crossover scales both parents' occupancies by $\kappa=0.5$, superimposes their effective atoms in the common fixed cell, perturbs positions and species weights, and prunes pairs closer than 1~\AA{} before EAT relaxation. No separate mutation offspring are enabled in this configuration.

For Fig.~\ref{fig:ptycho}, spatial views use a central $280\times280$ pixel display crop obtained by removing 20 measured pixels ($3.05128$~\AA{}) from each lateral edge. Structure views retain atomic centers within this footprint without changing their heights. Simulated projections are computed from the full optimized structures using Eq.~\eqref{eq:wse2_forward_image}, summed over the 12 measured depth slices, and then cropped laterally; all experimental and simulated projections share the same intensity scale. The reported projection correlations are Pearson coefficients over the $280\times280$ displayed pixels. The experimental depth view is a single isosurface of the normalized measured field at 0.20, with no fitted-blob replacement or depth exaggeration. Opacity decreases with distance along the viewing axis to expose foreground contours. 

In all applications, elemental reference energies are computed with the same FAIRChem model from elemental structures obtained from the Materials Project database~\cite{materialsproject1,materialsproject2}.

\subsection{Conventional GA baseline for ptychographic reconstruction}
\label{sec:methods:ptycho:conventional}

The conventional $\mathrm{WSe_2}$ baseline uses fully occupied W and Se atoms and the same TIFF crop, normalization, Gaussian image renderer, fixed periodic cell, formation-energy model, and image, composition, and sample-box penalties as HEAT. Its objective is Eq.~\eqref{eq:wse2_total_objective} evaluated for hard atoms with $\beta=0$. Both searches use populations of 50 members, with 5 elites, 13 random immigrants, and 32 crossover children per generation, and parent tournaments of sizes 4 and 2.

Random members are generated from the same intensity-derived cylindrical candidate pool described above. Locations are independently occupied with a probability chosen to give a mean initial population of 1500 atoms before overlap repair; occupied locations receive W or Se labels with probabilities $1/3$ and $2/3$. Crossover selects opposite sides of a randomly oriented plane from the two parents in their common experimental coordinate frame, without translating or aligning the parents. Each crossover child undergoes one mutation with probability 0.30, choosing rattle, relocation, insertion, or deletion with conditional probabilities 0.35, 0.25, 0.20, and 0.20, respectively. Rattle displaces 20\% of the atoms with a Gaussian standard deviation of 0.25~\AA{}. Proposed structures are repaired using a 1~\AA{} minimum separation.

Each non-elite candidate undergoes position-only projected-BFGS relaxation of the shared physical objective, with species and population fixed during local relaxation. The iteration limit, projected-gradient tolerance, and cost-plateau criteria match HEAT's final positional stage. The discrete baseline uses no soft image prefit, fractional-species optimization, syntropization, or rounding. Thus the two methods share the image-derived candidate pool but differ in initialization: HEAT performs a soft image prefit followed by effective-atom relaxation, whereas the conventional GA baseline directly initializes hard atoms.


\section*{Data Availability}
The datasets generated and/or analysed during the current study are available at https://github.com/jt577/HEAT-data. 

\section*{Code Availability}
The underlying code for this study is available at https://github.com/jt577/HEAT-data.

\bibliography{refs}

\section{Acknowledgements}
J.T., Z.S., D.M., and T.A. acknowledge support from the Center for Alkaline-Based Energy Solutions (CABES), part of the Energy Frontier Research Center (EFRC) program supported by the U.S. Department of Energy under grant DE-SC-0019445. C.-H.L. acknowledges support from the Eric and Wendy Schmidt AI in Science Postdoctoral Fellowship, a program of Schmidt Sciences. C.M. acknowledges support from the US National Science Foundation (NSF) under award PHY-1549132, the Center for Bright Beams (CBB). S.K. acknowledges support from Army Research Office grant W911NF-25-1-0256.

\section{Author Contributions}
J.T. conceived HEAT, developed all methods, performed all optimizations, and edited the manuscript. C.-H.L. proposed the $\mathrm{WSe_2}$ reconstruction application and provided PtyRAD and the experimental data. C.M. investigated early approaches to the image penalty. S.K. provided experimental and simulated ptychography data for other systems to test and improve the method. Z.S. contributed to scientific discussions. D.M., S.K., and T.A. originally proposed combining EAT with ptychography data. D.M. and T.A. supervised the study. T.A. edited the manuscript. All authors reviewed the manuscript.

\section{Competing Interests}
All authors declare no financial or non-financial competing interests.

\end{document}